\documentclass[twocolumn]{aastex631}
\usepackage{CJK}
\usepackage{framed}
\usepackage{multirow}
\usepackage{makecell}
\usepackage{amsmath, bm}
\usepackage{graphicx}
\usepackage{amssymb}
\usepackage{amsmath}
\usepackage{times}
\usepackage{subfigure}
\usepackage[T1]{fontenc}

\newcommand{\Msun}{\>{\rm M_\odot}}
\newcommand{\Mstar}{\>{\rm M_\star}}
\newcommand{\Mag}{\>{\rm mag}}

\newcommand{\Mpc}{{\rm \ {Mpc}}}

\newcommand{\kpc}{{\rm \ {kpc}}}

\received{}
\revised{}
\accepted{}

\submitjournal{ApJ}

\begin{document}
\begin{CJK*}{UTF8}{gkai}

\shorttitle{Satellite plane structures in TNG50}
\shortauthors{Hu and Tang}

\title[Satellite plane structures in TNG50]{Study of Satellite Plane Structure Characteristics Based on TNG50 Simulations: A Comparative Analysis from Plane to Non-Plane Structures}

\author{Caiyu Hu (胡才宇)}
\affiliation{School of Physics and Astronomy, China West Normal University, ShiDa Road 1, 637002, Nanchong, China}

\author[0000-0001-6395-2808]{Lin Tang (唐林)}
\affiliation{School of Physics and Astronomy, China West Normal University, ShiDa Road 1, 637002, Nanchong, China}
\affiliation{CSST Science Center for the Guangdong-Hongkong-Macau Greater Bay Area, DaXue Road 2, 519082, Zhuhai, China}

\correspondingauthor{Lin Tang}
\email{tanglin23@cwnu.edu.cn}

\begin{abstract}

In recent years, multiple plane structures of satellite galaxies have been identified in the nearby universe, although their formation mechanisms remain unclear. 
In this work, we employ the TNG50-1 numerical simulation to classify satellite systems into plane and non-plane structures, based on their geometric and dynamical properties. 
We focus on comparing the characteristics of these plane and non-plane structures.
The plane structures in TNG50-1 exhibit a mean height of $5.24 \kpc$, with most of them found in galaxy groups with intermediate halo virial masses within the narrow range of $\rm 10^{11.5}$ to $10^{12.5}\ \rm M_{\odot}$​. 
{Statistical analyses reveal that plane structures of satellite galaxies constitute approximately 11.30\% in TNG50-1, with this proportion increasing to 27.11\% in TNG100-1, aligning closely with previous observations. 
Additionally, central galaxies in clusters and groups hosting co-rotating plane structures are intermediate massive and slightly metal-poorer than those in non-plane structures.
Significant difference are found between in-plane and out-of-plane satellite galaxies, suggesting that in-plane satellites exhibit slightly longer formation times, and more active interstellar matter cycles.
The satellites within these plane structures in TNG50-1 exhibit similar radial distributions with observations, but are fainter and more massive than those in observational plane structures, due to the over- or under-estimation of galaxy properties in simulations.
Our analysis also shows that the satellite plane structures might be effected by some low- or high-mass galaxies temporarily entered the plane structures due to the gravitational potential of the clusters and groups after the plane structures had formed.}  

\end{abstract}

   
\section{Introduction}           
\label{sect:intro}

Numerous astronomical observations have demonstrated that the Lambda Cold Dark Matter ($\Lambda$CDM) cosmological model provides a robust framework for explaining the universe, particularly regarding large-scale structure and density distribution \citep[e.g.,][]{Bahcall1999}.
However, challenges persist at smaller scales. 
For instance, notable discrepancies remain between theoretical predictions and observational results regarding the distribution and formation models of dwarf galaxies \citep[e.g.,][]{Guo2011, McConnachie2012}. 
A particularly challenging issue is the planar arrangement of satellite galaxies.

The term ``plane structure of satellite galaxies'' refers to the planar distribution of satellite galaxies around their central galaxy within groups and clusters \cite[see][for a review]{Pawlowski2018}.
This phenomenon was first observed in the Milky Way (MW) by \cite{Lynden-Bell1976} and \cite{Kunkel1976}, and is referred to as the Vast Polar Structure (VPOS). 
The initially discovered plane in the Milky Way comprises the 11 classical brightest satellite galaxies, which move coherently within the same plane, with velocity vectors perpendicular to the plane's normal vector, indicating co-rotation.
Utilizing high-resolution simulations of the Milky Way's dark halo formation in the $\Lambda$CDM universe, \cite{Stoehr2002}  obtained kinematic characteristics of these 11 satellite galaxies that were consistent with observations. 
\cite{Kroupa2005} subsequently found that the plane of dwarf satellite galaxies in the Milky Way is nearly perpendicular to the galaxy's disc.
With the discovery of new Milky Way satellites such as Aquarius II, Crater II, and Antlia II \citep[e.g.,][]{Torrealba2016a, Torrealba2016b, Torrealba2019}, the study of the Milky Way's massive subhalo has progressed significantly.
Based on Gaia Early Data Release 3 (EDR3), \cite{Li2021} and \cite{Battaglia2022} explored the orbital properties of approximately 50 of the Milky Way's satellite galaxies, finding that nearly half lie on a plane. 
\cite{Xu2023} reported that in the TNG50-1 simulation, the 14 brightest satellite galaxies in Milky Way-like systems form an anisotropic plane almost perpendicular to the central stellar disc, consistent with observations.

Beyond the Milky Way, other plane structures of satellite galaxies have been identified in the nearby universe \cite[see][for a review]{Muller2023}. 
For instance, numerous studies \citep[e.g.,][]{Koch2006,  McConnachie2006, Metz2007, Conn2013} have found that 15 out of 17 satellite galaxies of M31 are aligned in a plane, known as the Great Plane of Andromeda (GPoA). 
Subsequent research by \cite{Ibata2013} demonstrated that only 13 of these 15 satellite galaxies exhibit coherent motion. 
Additionally, \cite{Shaya2013} identified another plane formed by some of M31's satellites, the existence of which has been confirmed by other studies \citep[e.g.,][]{Santos2020}. 
\cite{Tully2015} discovered the plane structures of satellite galaxies near Centaurus A (Cen A), finding that 27 out of 29 of Cen A's satellite galaxies form two thin planes, referred to as the Centaurus A Satellite Plane (CASP) structures. 
However, $\Lambda$CDM cosmological simulations predict that fewer than 1\% of galaxy groups and clusters exhibit double plane structures similar to those of Cen A \citep[e.g.,][]{Muller2018, Muller2021a}.
A plane structure of satellite galaxies has also been identified near NGC 253 \citep[e.g.,][]{Martinez-Delgado2021,Mutlu-Pakdil2024}.  
As the number of identified satellite galaxies increased, a new satellite plane of NGC 253 was discovered, exhibiting a larger minor-to-major axis ratio compared to the original plane formed by 5 satellites \citep[e.g.,][]{Martinez-Delgado2024}. 
Moreover, plane structures of satellite galaxies have been identified in other local groups, such as M81 \citep{Chiboucas2013}, NGC 4490 \citep{Pawlowski2024}, and NGC 2750 \citep{Paudel2021}, as well as in galaxy surveys like the Sloan Digital Sky Survey \citep{Phillips2015} and the Mass Assembly of early-Type GaLAxies with their fine Structures (MATLAS) survey \cite{Heesters2021}.

Significant focus has been placed on the formation mechanisms of satellite galaxy planes. 
It has been proposed that these planes form through the accretion of satellites along cosmic filaments or via tight group accretion influenced by the host galaxy \citep[e.g.,][]{Pawlowski2015, Pawlowski2021, Forster2022, Sato2024}.
\cite{Lynden-Bell1995} suggested that the planes of satellite galaxies result from the host galaxy's accretion, which imparts identical  angular momentum and orbital direction to the dwarf galaxies.
Using cosmological $N$-body simulations to study M31, \cite{Buck2015} discovered large dark matter filaments around galaxies in the high-redshift universe, providing evidence for accretion during the plane formation process.
Based on EDR3 and Hubble Space Telescope (HST) data, \cite{Julio2024} inferred that some recently discovered dwarf satellite galaxies near the Milky Way would be accreted into the Milky Way's halo, as predicted by \cite{Torrealba2016a}.
Additionally, it has been proposed that the merger of two galaxies could result in the remnants of the smaller galaxy's satellites forming a satellite galaxy plane \citep[e.g.,][]{Kanehisa2023}. 
However, such large-scale mergers might disrupt the original satellite plane, complicating the formation of new satellite galaxy planes \citep[e.g.,][]{Muller2021b}.
Contrary to the accretion and merger theories, some propose that tidal streams, formed through repeated mergers and tidal disruptions, eventually collapse to form tidal dwarf galaxies (TDGs), which subsequently form satellite galaxy planes under the influence of gravitational potential \citep[e.g.,][]{Wetzstein2007, Bilek2021, Banik2022}.
Furthermore, \cite{Xu2023} confirmed that the formation of satellite planes is influenced by the surrounding environment.

As \cite{Pawlowski2018} noted, the frequency of plane structures of satellite galaxies is fewer than 0.5\% in simulations, which is a significant small-scale challenge for the $\Lambda$CDM model. However, \cite{Phillips2015} and \cite{Heesters2021} found that approximately 10\% and 30\% of isolated host galaxies exhibit such satellite planes, suggesting that these satellite planes may be relatively common \citep{Cautun2015}. 
{It is believed that the existence of satellite planes is not unexpected in current models of galaxy formation \citep[e.g.,][]{Boylan-Kolchin2021, Sawala2023}, which supports the idea that satellite planes may not pose a significant challenge to the $\Lambda$CDM model.}

{\cite{Xu2023} identified a plane structure resembling the Milky Way  satellite plane in the TNG50-1 simulations by calculating the diameter-to-length ratio and thickness of the 14 brightest satellite galaxies. 
They attributed the formation of such MW satellite plane analogs to the peculiarities of their local environment. 
However, studying individual case provides limited insight into the statistical properties of satellite plane structures, making it challenging to fully understand their formation and evolution. 
Building on the work of \cite{Xu2023}, we leverage the large-scale IllustrisTNG cosmological simulation of TNG50-1 to identify a broader sample of satellite plane structures. 
We further analyze these analogs to investigate their structures and properties comprehensively.}
The structure of this paper is as follows: Section~\ref{sect:methodology} describes the methodology, including data sources, sample selection, structural morphological classification of satellite galaxy systems, and methods for identifying satellite planes.
 Section~\ref{sect:statistics} compares the properties of different structures in the simulations. 
 Section~\ref{sect:comparison} contrasts the TNG simulations with actual observations of satellite planes. 
 Finally, Section~\ref{sect:conclusion} presents a discussion and conclusions.

\section{Methodology}\label{sect:methodology}

\subsection{simulation}\label{sect:simulation}

In this work, the simulation we utilized is the TNG50 simulation of  IllustrisTNG suite \footnote{https://www.tng-project.org}.
Detailed descriptions of this database can be found in \cite{Pillepich2018a, Springel2018, Nelson2018, Naiman2018, Marinacci2018}. 
The IllustrisTNG simulations are an improvement of the previous Illustris simulations \citep[][]{Vogelsberger2014}, incorporating a revised active galactic nucleus (AGN) feedback model to ragulate star formation efficiency in massive galaxies \citep{Weinberger2017}, and a galactic wind model to inhibit efficient star formation in low- and intermediate-mass galaxies \citep{Pillepich2018b}.
The TNG50-1 simulation \citep{Nelson2019a, Nelson2019b, Pillepich2019} is the highest resolution version of the IllustrisTNG suite, with $2160^3$ dark matter and $2160^3$ gas particles in a cubic box of $(51.7 \ \rm Mpc)^3$.
The mass resolution is $8.5 \times 10^4 \ \rm M_\odot$ for baryon particles and $4.5 \times 10^5 \ \rm M_\odot$ for dark matter, while the spatial resolution is set as $72 \ \rm pc$.
The cosmological parameters are derived from Planck \citep{Planck2016}: $\Omega_m=0.3089$, $\Omega_{\Lambda}=0.6911$,  $\Omega_{b}=0.0486$,  $n_s=0.9667$, $\sigma_8=0.8159$, and $h=0.6774$.
Dark matter halos are identified using the Friends-of-Friends (FoF) algorithm \citep{Davis1985}, while subhalos are identified as overdense, gravitationally bound substructures using the SUBFIND algorithm \citep{Springel2001, Dolag2009}.

FoF groups are selected with $\rm M_{200}  >  10^9\Msun$, where $\rm M_{200}$ represents the mass within a radius where the average density is 200 times the critical density of the Universe.
Subhalos are included if they have a total stellar mass $\Mstar>  10^5 \Msun$. 
{In many theoretical studies, subhalos are typically selected to  include more than 100 stellar particles, which corresponds to $\rm \Mstar > 10^7\Msun$ in the TNG50-1 simulation.
However, most observed satellite galaxies in plane structures, such as those in the Milky Way and M31, have stellar masses below $10^7\ \Msun$.
To enable a meaningful comparison with observations, we adopt a lower stellar mass threshold of $\Mstar > 10^5\Msun$ in the procedure of satellite plane finding, consistent with the selection criteria used in \cite{Xu2023}, which balances simulation resolution limitations with observational completeness.
We acknowledge that subhalos with $\rm 10^5\Msun < \Mstar < 10^7 \Msun$ are represented by fewer than 100 stellar particles in the simulation, leading to weaker statistical significance for results derived from this mass range. 
So we remove the satellite galaxies with $\Mstar < 10^7 \Msun$ in Section~\ref{satellites} and ~\ref{satellite in and out} to reduce the effects of the simulation mass resolution as much as possible.
Overall, The selection criterion of $\Mstar > 10^5 \Msun$ allows us to include a larger number of potential samples, enhancing the statistical robustness of the analysis.
We emphasize that conclusions drawn for subhalos within the range $\rm 10^5\Msun < \Mstar < 10^7\Msun$ should be interpreted with caution due to resolution limitations.} 
Only groups containing more than 6 subhalos are considered, excluding the centrals, which are the most massive subhalos within the groups.
{The N > 6 criterion is based on observations, as the smallest number of satellites in observed satellite plane structures is approximately 5 \citep[e.g.,][]{Karachentsev2003, Martinez-Delgado2021}. 
The addition of one more satellite ensures statistical reliability across our definitions.}
The snapshot used for analysis is taken at redshift $z=0$ (snap number = 99).
The final sample comprises 699 FoF groups.

\subsection{Definition of satellite planes}

We compare previous methods for identifying satellite planes  \citep[e.g.,][]{Gillet2015, Pawlowski2024} and find that most of the random projection ellipsoid fitting approaches do not account for the mass of satellite galaxies, focusing instead solely on fitting the spatial positions of the member galaxies and considering motion coherence. 
This can lead to significant issues, as low-mass galaxies located outside the satellite plane and at large distances can substantially affect the satellite plane's orientation. 
This results in a deviation between the primary satellite plane and the true satellite plane. 
To mitigate the impact of mass distribution on satellite plane identification, it is crucial to incorporate mass weighting into the analysis. 

Nearly half of the satellite galaxies in the Milky Way and M31 have been observed to align within plane structures.
However, if the member galaxies not residing on the satellite plane have substantial mass, are numerous, or are located far from the satellite plane structures, this uneven distribution can significantly   affect the calculation of shape parameters \citep[e.g.,][]{Libeskind2016, Pawlowski2017,Gong2019, Wang2021}, such as the eigenvalues and eigenvectors of the inertia tensor. 
As with previous random projection ellipsoid fitting methods, this  uneven distribution can increase the angular error between the fitted plane and the true plane.

To accurately determine geometric and dynamical properties of the plane structures of satellite galaxies, we develop a method based on the Random Sample Consensus (RANSAC) algorithm.
The RANSAC algorithm \citep{Raguram2013} enables robust fitting by randomly selecting inliers and iteratively managing noise and outliers in the data. 
By selecting the model with the highest number of inliers as the final result, this approach allows us to obtain relatively accurate fitting even in the presence of noise and outliers.
This process involves excluding satellite galaxies farther from the principal plane than a distance $D_0$​, enabling a more precise determination of the principal axis direction of the fitted ellipsoid. 
Here, $D_0$ represents the vertical distance from the plane using in the RANSAC algorithm.

Given the influence of gravitational potential energy and the conservation of momentum in kinematics, satellite members within a plane structure must maintain consistent motion within the same plane. 
As a result, their rotational radii are relatively small, causing these co-rotating satellites to be tightly aligned with the central galaxy. 
Observational comparisons reveal that the physical sizes of plane structures typically range from 10 to 100 $\kpc$. 
Beyond this distance, the satellite distributions become more random, and no plane structures are observed. 

With careful consideration, we utilize a radius of $100 \kpc$  to filter satellites. 
Satellites located beyond this distance from the central galaxies are then removed.
This step helps avoid the influence of distant satellites on the identification and analysis of satellite planes,  as previously discussed.
The remaining satellites are then subjected to inertia tensor calculations and corrections.
To ensure the accuracy of our calculations, we select galaxy clusters that contain more than 6 satellite galaxies within $100 \kpc$ of the central galaxies. 
This criterion results in a sample of 347 out of the 699 groups in the principal sample, with the remaining groups classified as non-plane structures.
The procedure we use to define plane structures of these 347 groups is introduced below.
\begin{table*}
\begin{center}
\caption{Three types of satellite structures and selection criteria. 
$N$ represent the number of satellite galaxies with $100\kpc$ of the central galaxy.}\label{sample}

 \begin{tabular}{clclclc}
  \hline\noalign{\smallskip}
Structure Type &  Number  & $S_{mean}$  & $H_{mean}[\kpc]$ & $K_{mean}$ &  $N$ \\
  \hline\noalign{\smallskip}
    thin plane   & 72  & [0.75, 1]    & [0, 10]            & [4.5, $+\infty$] & [6, $+\infty$] \\ 
   thick plane   & 7   & [0.75, 1]    & [10, 15]           & [6, $+\infty$]    & [6, $+\infty$] \\        
pseudo-plane & 9   & [0.75, 1]    & (10, 15]           & [4.5, 6)             & [6, $+\infty$] \\
non-plane     & {611} & [0.5, 0.75) & [15, $+\infty$]  & [0, 4.5]            & [0, 6)      \\
  \noalign{\smallskip}\hline
\end{tabular}
\end{center}
\end{table*}
\begin{enumerate} 
\item Inertia Tensor Calculation: We begin by computing the inertia tensor $T$ of the galaxy group using the satellite members within $100\kpc$ centred at central galaxy, following the method of \cite{Xu2023}.
The inertia tensor is calculated as follows:
\begin{equation}
T=\sum_i M_{i}\bf({r}_i-r_0)(r_i-r_0)^T
\label{eq:1}
\end{equation}
where $M_i, r_i, r_0$ represent the total mass of the satellite galaxies, the position of the satellite galaxies, and the position of the central galaxies, respectively.
The three eigenvectors of the inertia tensor correspond to the directions of the fitted ellipsoid's major, semi-major, and minor axes. 
The original plane of the satellite structure is defined by the major  and semi-major axes.

\item RANSAC algorithm: Then, we determine the principal membership of the plane structures by applying the RANSAC algorithm to the original plane defined in the first step.
Following careful analysis, we use a threshold distance $D_0 = 50 \kpc$ to filter the satellites. 
Satellites located beyond $50 \kpc$ are excluded, refining the identification of the satellite membership.
To accurately determine the membership of the satellite planes, we perform multiple iterations, reducing the threshold distance with $D_{i+1} = 0.9 \times D_{i}$, where $i$ represents the iteration number. 
This filtering process is repeated up to 20 times or until the number of satellites falls below 6 or less than half of the original count.
Note that the plane used in RANSAC algorithm may increasingly approach the principal plane as the process is repeated.
The precise shape of the satellite planes can be computed using the inertia tensor of the principal membership of the satellite planes determined by the RANSAC algorithm.

\item Kinematic Analysis: We then project the velocities and coordinates of the satellite galaxies onto the system's principal plane and calculate the spin $S_i$ of each satellite galaxies selected in first step:
\begin{equation}
S_{i}=\bf({r}_i-r_0)\times(v_i-v_0)
\label{eq:2}
\end{equation}
where $\bf v_i$ and $\bf v_0$ are the velocities of the satellite galaxies and central galaxy, respectively.   
The sign of $S_i$ indicates the direction of satellite rotation. 
By counting the positive and negative direction ($N_{+}$ and $N_{-}$), we calculate $K$, the maximum of these two numbers, and divide it by the total number of satellites in the plane to obtain the scaled system spin $S$. 
A larger $K$ and $S$ indicate more satellites co-rotating and stronger kinematic coherence among satellite galaxies.
$S=1$ means that all the satellite galaxies on the plane structures rotate with a same direction, i.e., $100\%$ co-rotation.

\item Structure Height Calculation: We next calculate the root-mean-square (RMS) height $H$ to define the plane thickness, similar to the approach used in \cite{Samuel2021}.
A smaller $H$ suggests a thinner plane structure.

\item Repetition and Classification: Finally, we repeat the procedures 50 times to obtain the mean values $S_{mean}$, $H_{mean}$, and $K_{mean}$ for $S$, $H$, and $K$.
Based on these parameterized geometric and dynamical properties, we classify the satellite structures into the following three types: 
\begin{enumerate} 
\item{\it Plane Structures}: $H_{mean} < 10\kpc$, $S_{mean} > 0.75$, and $K_{mean} > 4.5$, or $10 < H_{mean} < 15\kpc$, $S_{mean} > 0.75$, and $K_{mean} > 6$;  
\item{\it Pseudo-Plane Structures}: $10\kpc < H_{mean} < 15\kpc$, $S_{mean}> 0.75$, and $4.5 < K_{mean} < 6$; 
\item{\it Non-Plane Structures}: $H_{mean} > 15\kpc$, or $S_{mean} < 0.75$, or $K_{mean} < 4.5$.
\end{enumerate}
Additionally, there are {352} non-plane structures defined by having less than 6 satellite galaxies within $100 \kpc$ of the central galaxies. 
\end{enumerate}

\begin{figure}
   \centering
   \includegraphics[width = 0.5\textwidth]{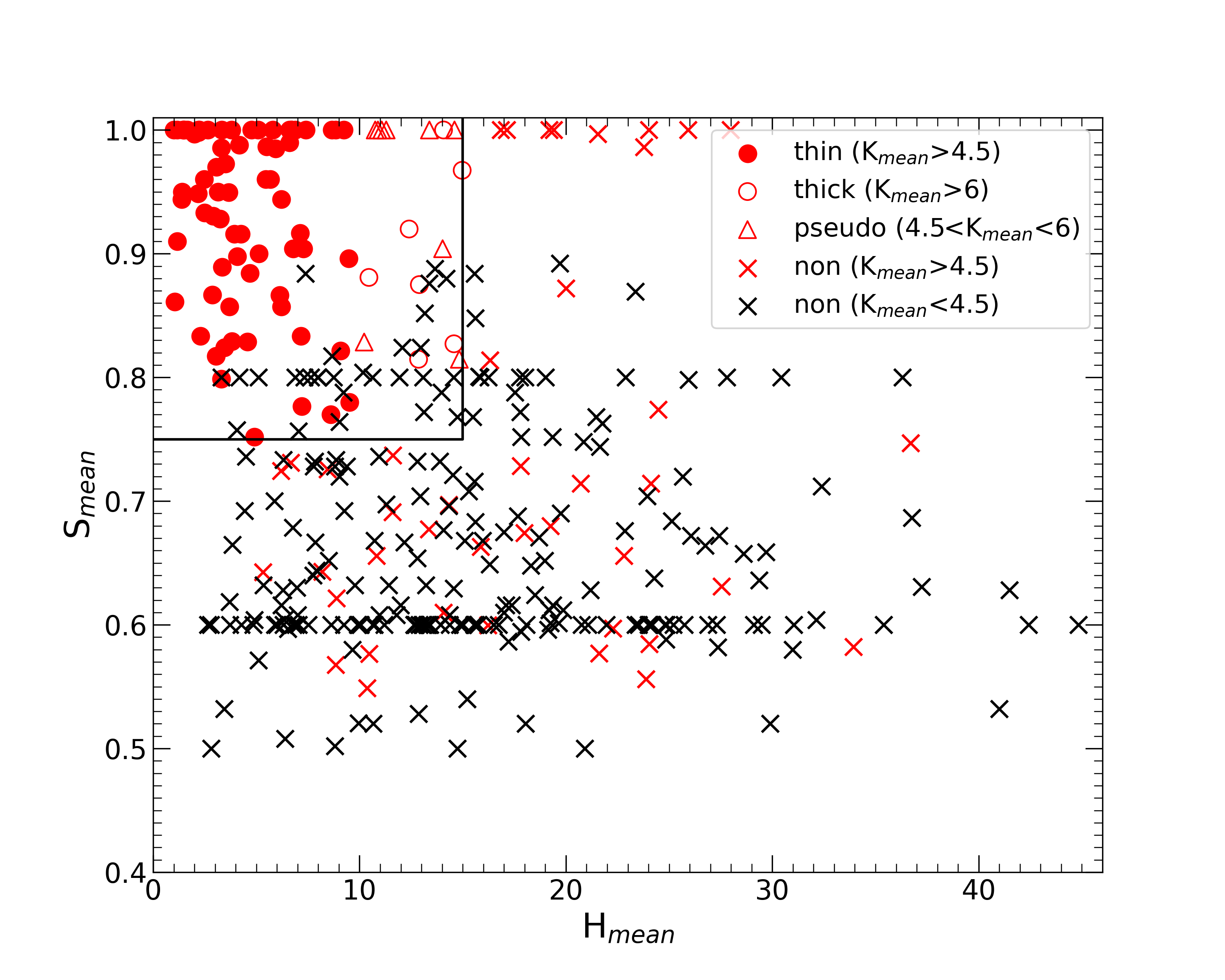}
   \caption{
   Distribution of $S_{mean}$ and $H_{mean}$ for satellite planes.
   The vertical and horizontal solid lines in the figure represent $S_{mean}=15\kpc$ and $S_{mean}=0.75$, respectively.
   The red symbols correspond to $K_{mean}> 4.5$.
   Based on this figure, the sample can be classified into four structures: {\it thin, thick, pseudo-plane, and non-plane}.
   Planes and pseudo planes are represented by all red data points within the box defined by the solid lines of the figure ($S_{mean}<15\kpc$, $S_{mean}>0.75$, and $K_{mean}>4.5$). 
   Non-planes are indicated by the data points either below the horizontal line ($S_{mean}<0.75$), to the right of the vertical line ($S_{mean}>15\kpc$), or within the box shaded black ($K_{mean}<4.5$).
   Thin planes are structures with $S_{mean}<10\kpc$.
   Thick planes are characterized by $10\kpc<S_{mean}<15\kpc$ and $K_{mean}>6$, while pseudo-planes have $10\kpc<S_{mean}<15\kpc$ and $4.5<K_{mean}<6$. 
   }
   \label{H_S_K}
\end{figure}
\begin{figure*}
   \centering
   \includegraphics[width=\textwidth]{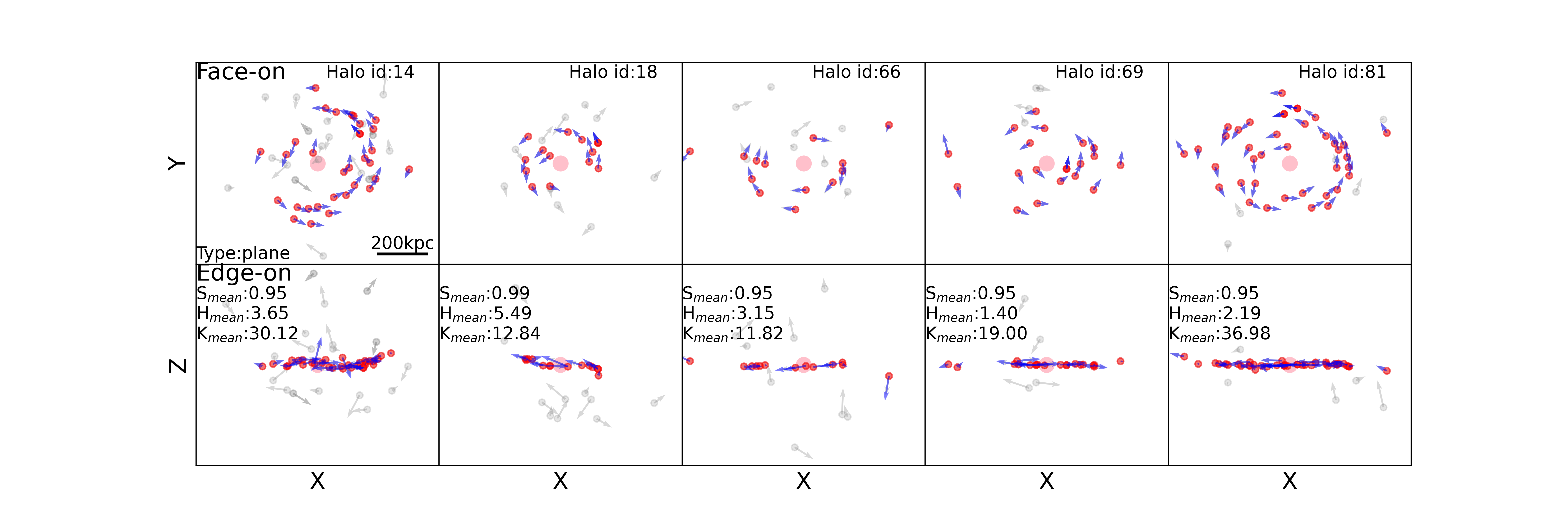}
   \includegraphics[width=\textwidth]{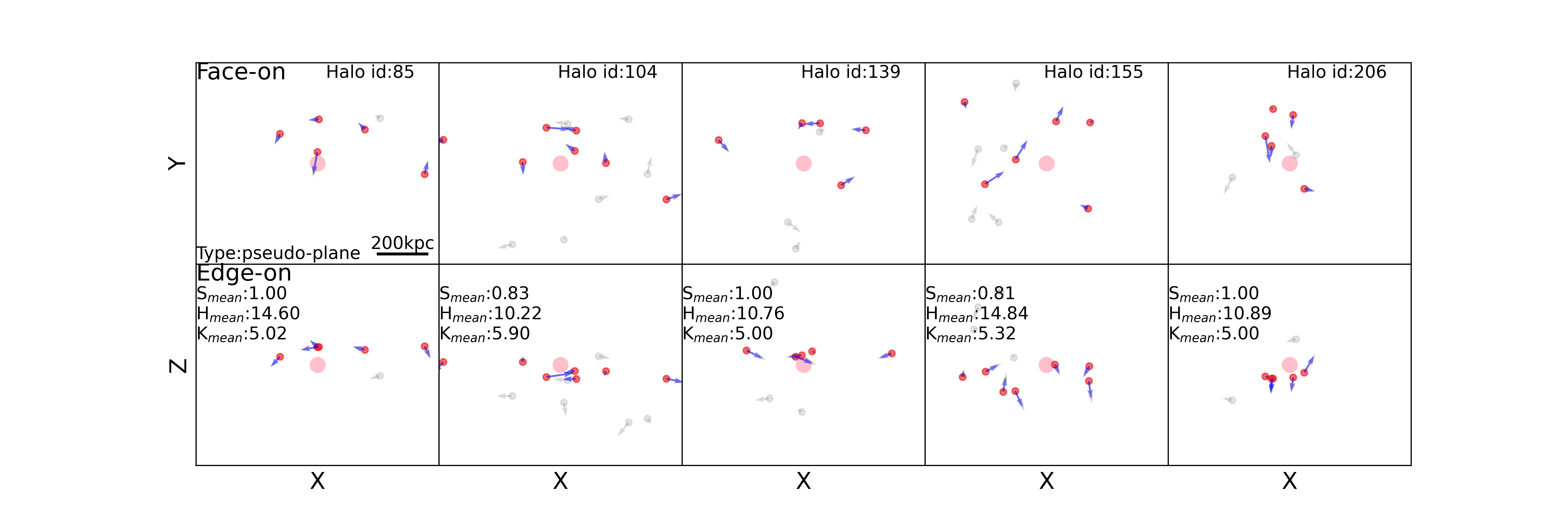}
   \includegraphics[width=\textwidth]{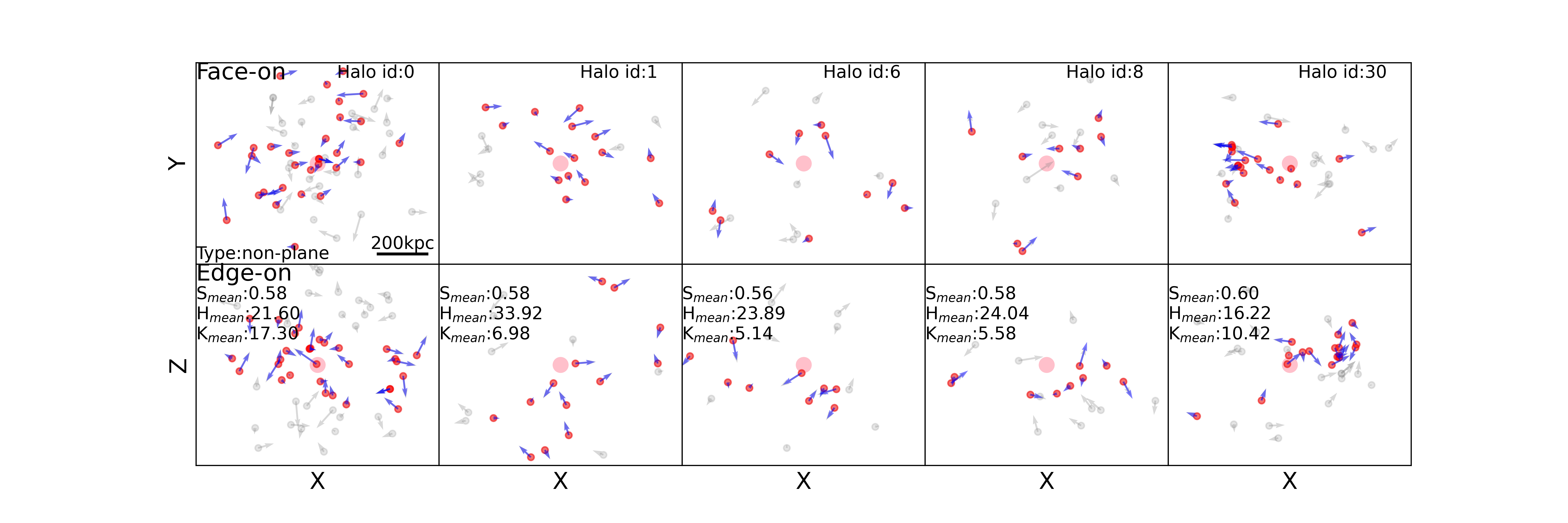}
   \caption{
   Face-on and edge-on projection views of the three types of satellite structures, with five examples selected for each structure. 
The panels are arranged from top to bottom for plane, pseudo-plane, and non-plane structures, respectively.
The central galaxies are represented by large pink circles. {The satellite galaxies within the plane structures are marked by red dots, and blue arrows indicate the projected velocity vectors, while those outside the plane structures are marked by gray dots and arrows.} 
The face-on views correspond to the primary plane of flattening, consisting of the longest and second-longest axes. 
The edge-on views are defined by the planes formed by the longest and shortest axes.
Halo id, $S_{mean}$, as well as $H_{mean}$, and $K_{mean}$ values, are also labeled in the panels. 
}
   \label{sample}
\end{figure*}

$S_{mean}> 0.75$ indicates that more than 75\% of the projected velocities of satellite galaxies in the planes point in the same rotation direction.
$H_{mean}$ quantifies the thickness of the structures, with $10$ and $15\kpc$ selected as the best thresholds to differentiate between plane, pseudo-plane, and non-plane structures based on careful evaluation.
$K_{mean} > 4.5$ signifies that, in multiple grogram runs, the majority of $K$ values are greater than 5, with some less than 5; this threshold of 5 is based on current observations \citep[e.g.,][]{Martinez-Delgado2024}.
Defining {\it pseudo-plane structures} using the criteria of $10\kpc < H_{mean} < 15\kpc$, $S_{mean}> 0.75$, and $K_{mean} > 4.5$ includes structures similar to those identified as marked planes with $K_{mean}$ values exceeding 6.
Consequently, we introduce an additional criterion of $10\kpc < H_{mean} < 15\kpc$, $S_{mean}> 0.75$, and $K_{mean} > 6$ to distinguish {\it thick plane structures}. 
For instance, the plane structure in $Halo395$ has $H_{mean} = 12.89\kpc$, $S_{mean} = 0.875$, and $K_{mean} = 13.6$.
The significantly smaller thickness of the plane structure in $Halo395$ compared to the $27\kpc$ reported by \cite{Xu2023} is due to our $H_{mean}$ being calculated based on satellites with the planes centred on the central galaxies, which differs sightly from  \cite{Samuel2021} and \cite{Xu2023}, where the calculations are  centred on the the geometric center of the satellites.
Structures with $H_{mean} < 10\kpc$ are classified as {\it thin}.
Table~\ref{sample} lists the numbers and selection criteria for the four types of structures in our samples.
Note that {\it plane structures} (including thin and thick) and {\it pseudo-plane structures} must satisfy all four selection criteria simultaneously, while {\it non-plane structures} only need to meet any one of these criteria.
Fig.~\ref{H_S_K} illustrates the distribution of $H_{mean}$ and $S_{mean}$.
By incorporating $K_{mean}$, we can classify our sample into the four types of structures as shown in Fig.~\ref{H_S_K}.
For the analysis, we will treat both {\it thin} and {\it thick plane structures} as unified {\it plane structures}.

Fig.~\ref{sample} illustrates the three types of structures, presented from top to bottom panels, with face-on (first row) and edge-on (second row) views of 2D coordinates and velocities. 
{Red and gray dots represent satellite galaxies within the plane and outside the plane structures, respectively.}
The face-on views (first row in the top panel) clearly show that the satellite galaxies within {\it plane structures} exhibit distinct co-rotation around the central galaxies. 
The edge-on views (second row in the top panel) reveal that most of these structures are very thin, with a mean thickness of $5.24\kpc$. 
These structures are primarily found in galaxy groups with intermediate halo viral masses with a narrow range from $\rm 10^{11.5}$ to $10^{12.5}\Msun$.
Thick plane structures (not shown in the figure) also have numerous sub-members rotating around the centers but appear significantly thicker in the edge-on view, with a mean thickness of $13.18\kpc$. 
Pseudo-plane structures resemble planes in the edge-on view (first row in the middle panel). 
However, the velocities of their sub-members are uncorrelated, displaying a dispersion distribution (second row in the middle panel). 
In contrast, the satellites in the non-plane structures neither form a plane nor co-rotate around the central galaxies in any directions  (bottom panel).

{Note that mass weighting is not directly considered in the RANSAC algorithm, as it has already been incorporated in the calculation of the inertia tensor. 
The RANSAC algorithm effectively removes distant satellite galaxies that deviate significantly from the plane, thereby reducing their impact on the accurate determination of the satellite plane during the inertia tensor calculation. 
Compared to previous methods, the RANSAC algorithm provides robust outlier rejection and iterative refinement, enhancing the accuracy of satellite plane definitions. 
For instance, unlike \cite{Xu2023}, which calculates the inertia tensor once, our approach involves multiple iterations to refine the plane. Additionally, the combination of RANSAC and inertia tensor calculations improves accuracy compared to the random projection method \citep[e.g.,][]{Pawlowski2024, Gillet2015} by reducing the influence of distant galaxies.}

\begin{figure*}
   \centering
   \includegraphics[width=\textwidth]{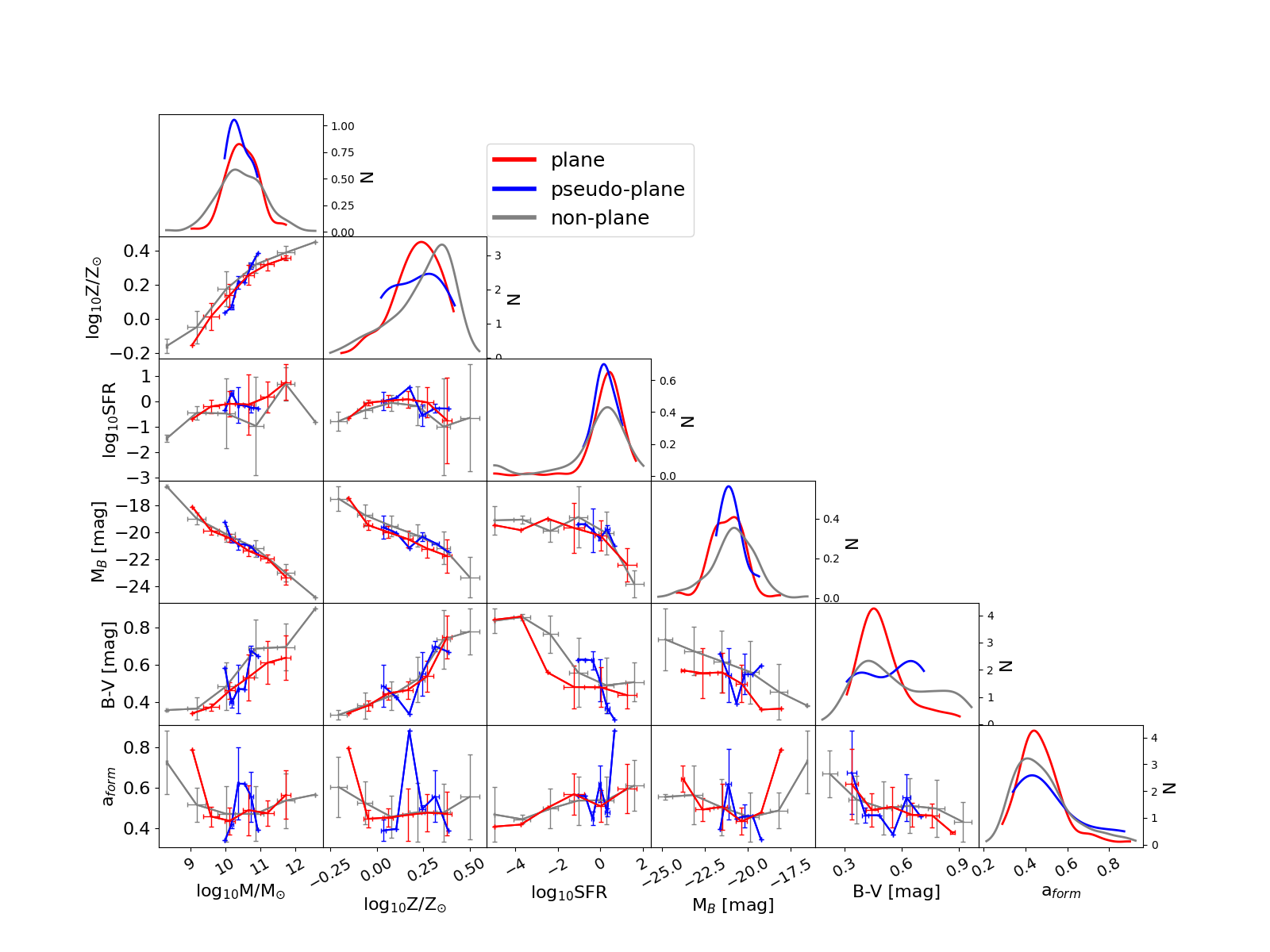}
   \caption
   {Cross-relations of central galaxies for stellar mass $\rm M_\star$, absolute magnitude in B band $\rm M_B$, stellar metallicity $\rm Z$, star formation rate $\rm SFR$, $\rm B-V$ color, and formation time of the located halo $a_{form}$; the number distributions of each properties.
   Error bars are calculated by the standard deviation. 
   Plane, pseudo-plane, and non-plane structures are marked by the red, blue, and grey lines, respectively.
   }
   \label{central_statistics}
\end{figure*}
\begin{figure*}
   \centering
   \includegraphics[width=\textwidth]{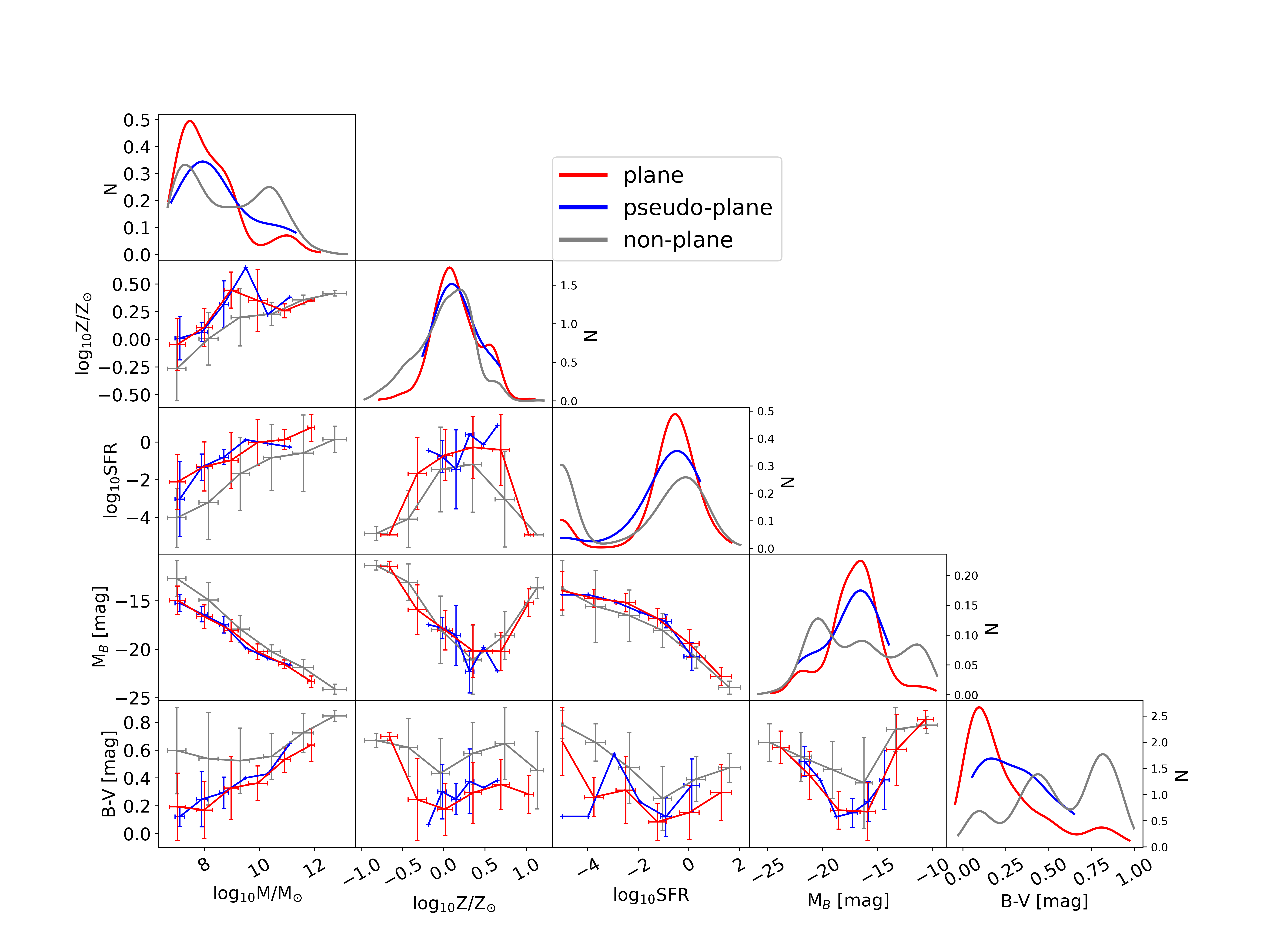}
   \caption
   {
   As Fig.~\ref{central_statistics}, but showing the property cross-relations of satellite galaxies within each structures. 
   Note that the formation time scale ($a_{form}$) is excluded from this analysis.
   }
   \label{satellite_statistics}
\end{figure*}
\begin{figure*}
   \centering
   \includegraphics[width=\textwidth]{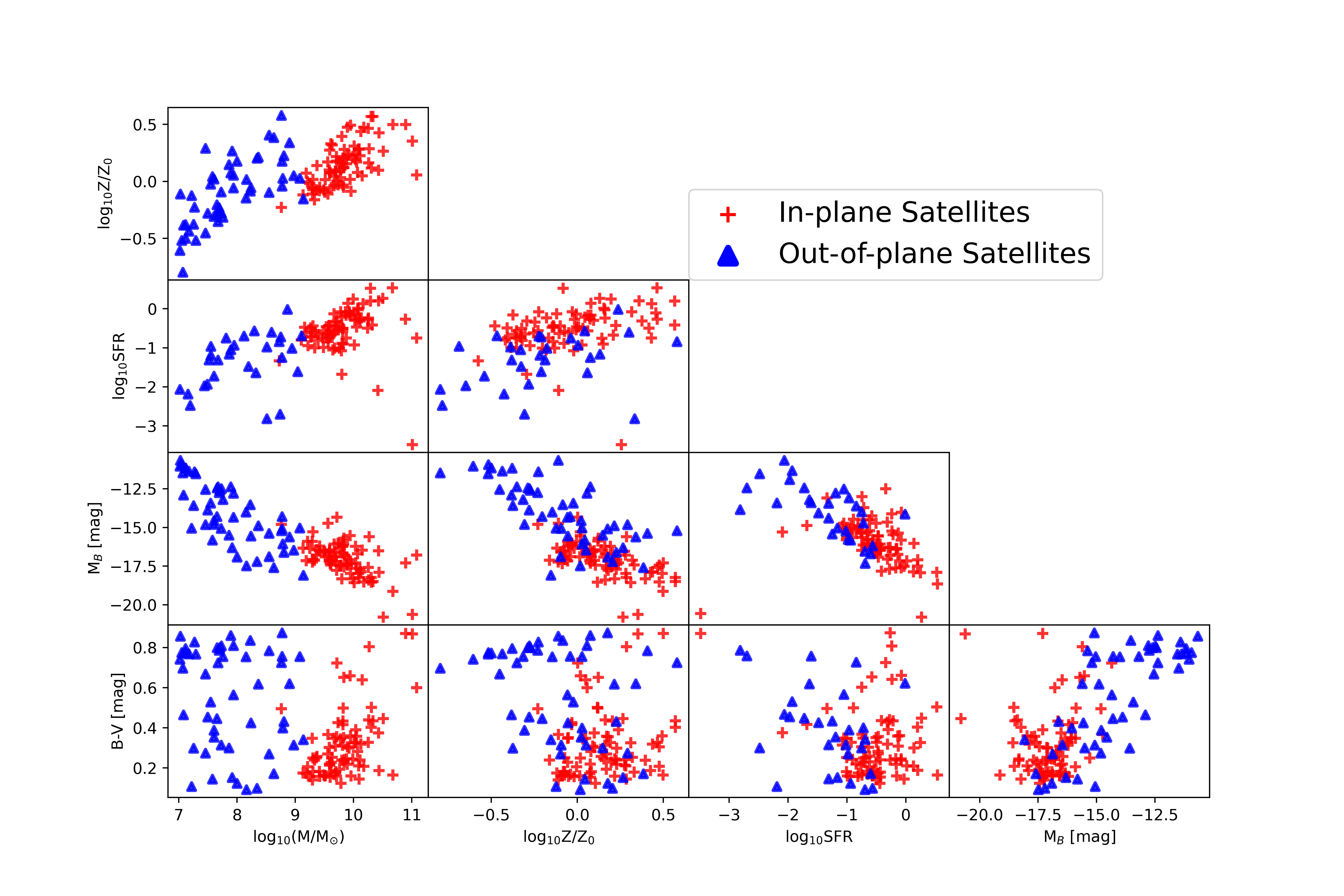}
   \caption
   {Cross-relations of satellite galaxies within planes (plus symbols) and outside of planes (triangle symbols) in plane structures.
   {The plots represent the average values of the properties of satellite galaxies within plane and outside of plane structures, in each groups.}
   }
   \label{satellite_in_out_disc}
\end{figure*}
\begin{figure*}
   \centering
   \includegraphics[width=\textwidth]{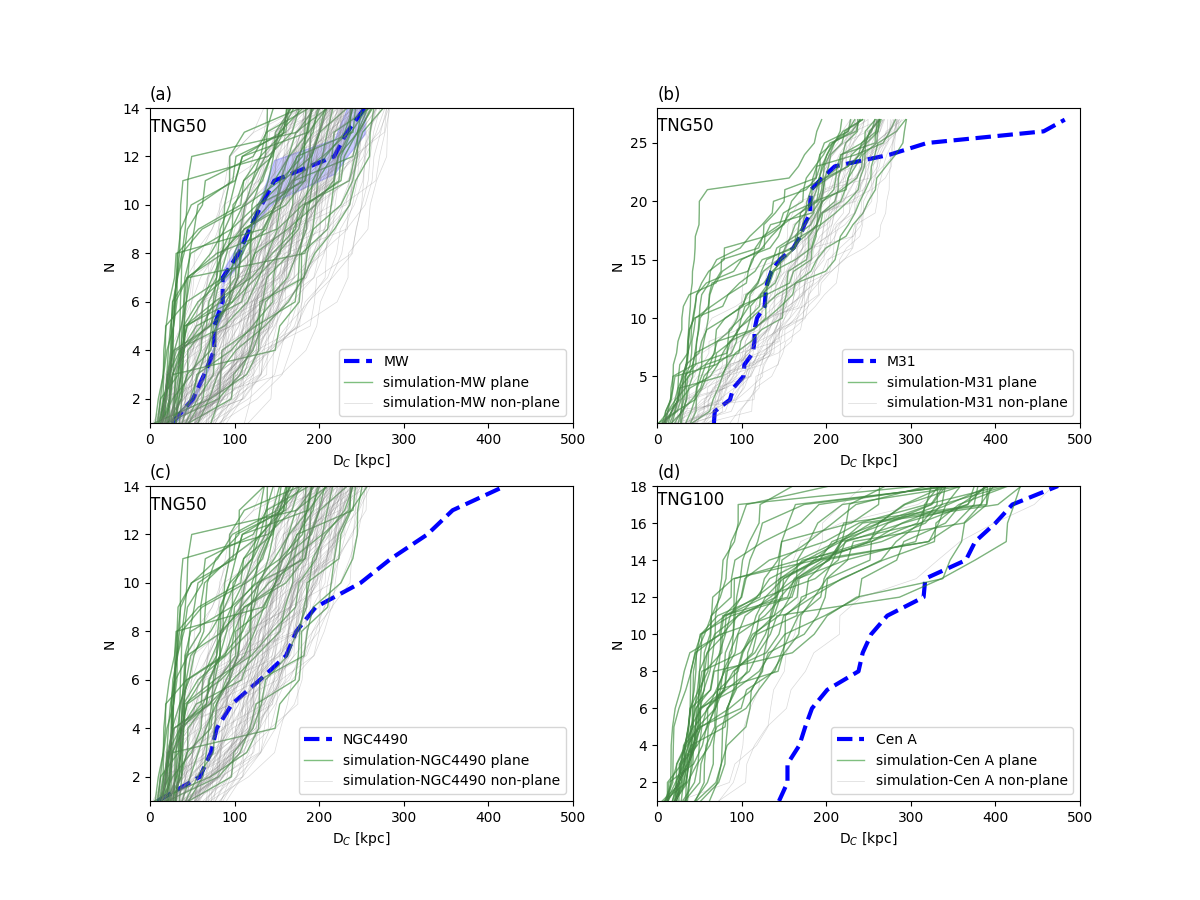}
   \caption
   {Radial distributions of satellite galaxies in MW (panel a), M31 (panel b), NGC 4490 (panel c), and Cen A (panel d) for simulated analogs (solid lines) and observational groups (blue long-dashed lines).
   The grey and green solid lines are the simulated analogs with and without plane structures, respectively.
   $D_c$ represents the distance from the central galaxies.
   The observational datas are taken from \cite{McConnachie2012, Torrealba2016a, Torrealba2019} for MW, \cite{Conn2012} for M31, \cite{Karachentsev2024} for NGC 4490, \cite{ Kanehisa2023} for Cen A.
   }
   \label{comp_1}
\end{figure*}
\begin{figure*}
   \centering
   \includegraphics[width=\textwidth]{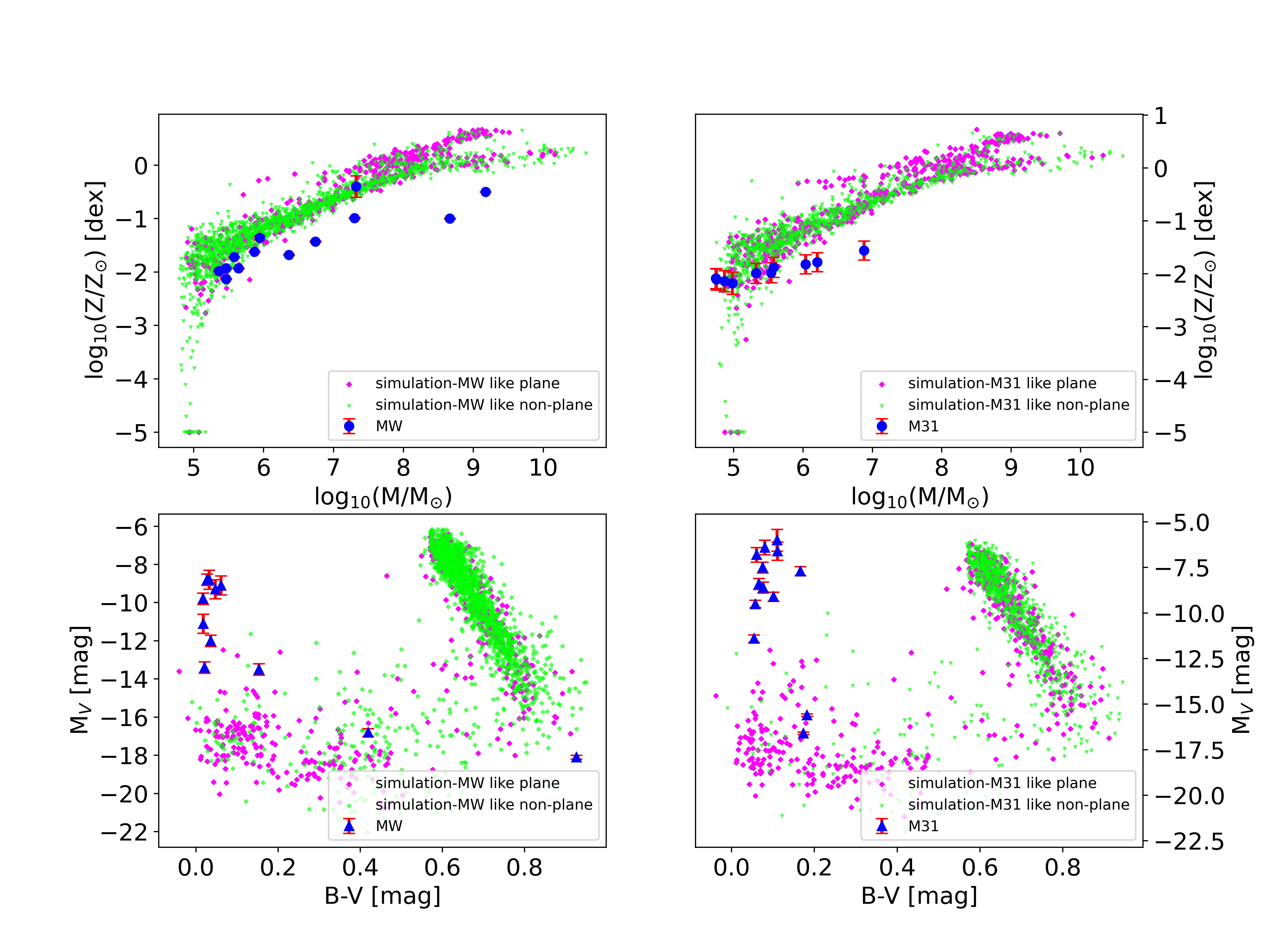}
   \caption
   {The metallicity -stellar mass (top panels) and magnitude - color (bottom panels) relationships of simulated analogs and observational groups for MW  (left panels) and M31 (right panels).
   The purple and green diamond symbols are the simulated analogs with and without plane structures, respectively, while triangle symbols are the observational results for MW and M31.
   The observational datas are taken from \cite{McConnachie2012, Torrealba2019} for MW, \cite{Conn2012, Santos2020, Savino2022, Jennings2023} for M31.
   }
   \label{comp_2}
\end{figure*}

\section{Properties of plane Structures}
\label{sect:statistics}
In this section, we first compare the properties of central galaxies associated with the three types of satellite structures (plane, pseudo-plane, and non-plane) to investigate the relationship between satellite plane structure and their central galaxies.
We then analyze the properties of satellite galaxies within these  structures.
Finally, we compare satellites located within and outside the satellite planes to identify the primary contributors to the formation of these plane structures. 
Specifically, we focus on several properties: galaxy stellar mass $\rm M_\star$, absolute magnitude in the B band $\rm M_B$, stellar metallicity $\rm Z$, star formation rate $\rm SFR$, $\rm B-V$ color, and the formation time of the halo $a_{form}$, which is defined as the cosmological scale factor at which the halo first attained half of its current mass.

\subsection{Property Cross-Relations of Central Galaxies}\label{central}

As shown in Fig.~\ref{central_statistics}, we present the cross-relations and number distributions for five different properties. 
Error bars represent the standard deviation. 
The number distributions in this figure clearly indicate that central galaxies with plane structures differ significantly from those without plane structures. 
{In the top panel of Fig.~\ref{central_statistics}, we find that central galaxies with satellite planes exhibit a more concentrated distribution compared to those without satellite planes. 
Approximately 80\% of central galaxies with satellite planes have stellar masses in the range of $10^{10}$ to $10^{11}\Msun$, whereas only about 50\% of central galaxies without satellite planes fall within this intermediate range. 
Moreover, central galaxies with satellite planes tend to occupy specific parameter ranges, including stellar metallicities of $\log_{10}(\rm Z/Z_{\odot})$ between 0 and 0.4, star formation rates ($\rm SFR$) from $\rm 10^{-2}$ to $\rm10^2\ M_{\odot}/yr$, B-band magnitudes $\rm M_B < -20 \ mag$, $\rm B-V$ colors range from $0.3$ to $\rm0.6\ mag$, and formation time scales ($\rm a_{form}$) between 0.3 and 0.6.
The stellar metallicity distribution of central galaxies with satellite planes also slightly higher than that of those without satellite planes.
These findings suggest that satellite plane formation is more likely to occur around central galaxies that are star-forming within and fall within intermediate ranges of stellar mass, metallicity, age and luminosity.
However, it is noted that there is still significant overlap in the properties of central galaxies with and without satellite planes, which makes directly distinguishing between the two populations challenging.} 

For a given stellar mass, central galaxies with plane structures are typically bluer, metal-poorer, and have higher star formation rates compared to their counterparts without satellite planes. 
At lower star formation rates, these plane structures still exhibit a blue color. 
Overall, central galaxies with satellite planes tend to be brighter, bluer, and have slightly higher metallicity compared to those in non-plane structures. 
The properties of central galaxies with pseudo-plane structures are more dispersed due to the small number of such structures, but their distributions are similar to those of plane structures, falling between plane and non-plane structures. 
This suggests that pseudo-plane structures may represent an intermediate state between plane and non-plane structures.

\subsection{The Satellite Properties of plane Structures}\label{satellites}

Due to the gravitational potential interactions between the satellite and central galaxies -- whether through mergers or accretion -- it is essential to consider the properties of each satellite galaxy within their respective dark matter halos.
In this section, similar to the analysis in Section.\ref{central}, we aggregate data for the satellite galaxies of each structural type and present the results as shown in Fig.~\ref{satellite_statistics}. 
{Note that we remove the satellite galaxies with $\Mstar<10^7\Msun$ to reduce the effects of simulation mass resolution in this section.} 

From the number distributions in Fig.~\ref{satellite_statistics}, it is evident that the satellite galaxies associated with plane structures exhibit more pronounced differences compared to those located in the non-plane structures than the central galaxies do. 
{The satellite galaxies in plane structures are primarily concentrated around $\Mstar\sim10^8\Msun$, while those in non-plane structures exhibit a more uniform distribution across stellar mass $\Mstar > 10^7\Msun$.} 
{Additionally, the satellite galaxies in plane structures are much brighter and bluer than those in non-plane structures, with almost all below $\rm-15\ mag$ in $\rm M_B$ and 0.25 $\rm mag$ in $\rm B-V$ color index.
However, number distributions of the metallicity and star formation rate seem to be similar for both satellite with plane and non-plane structures.
These two populations generally exhibit a number distribution peaked at $\rm\log(Z/Z_{\odot})\sim0$ and $\log\rm SFR\sim-0.5$.
For a given stellar mass, it is obvious that satellite galaxies in plane structures are bluer, brighter with higher metallicities and star formation rates than those in non-plane structures.} 
Similar to the analysis in the Section~\ref{central}, pseudo-planes  display distributions between plane and non-plane structures but are closer to those planes in both the number distributions or the property cross-relations.

As shown in the cross-relations images in Fig.~\ref{satellite_statistics}​, there are similarities between satellite galaxies in plane and non-plane structures for masses $\Msun > 10^{10}\Msun$. 
In these extreme high-mass ranges, the properties of satellite galaxies in plane structures closely resemble those in non-plane structures. 
We speculate that these satellites may not genuinely belong to the plane structures but instead appear within the planes by coincidence.  
Additionally, the mass resolution of simulated satellites could also affect the results at the low-mass end.

\subsection{Comparison of In-plane and Out-of-plane Satellites}\label{satellite in and out}

In Section~\ref{satellites}, we focus on satellites within the plane structures, which reveal significant differences from non-plane structures. 
However, when considering all satellites in plane structures, the differences with non-plane structures become less pronounced, particularly at the low or high mass end. 
This is likely due to the influence of out-of-plane satellites in plane structures. 
Satellite galaxies within the plane structures exhibit strong kinematic consistency, with nearly all rotating around the central galaxy, experiencing strong gravitational potential from the central galaxy, and displaying great kinematic correlation. 
These factors theoretically make them more suitable for studying the properties of satellite planes.
Therefore, further analysis of plane structures is necessary to explore the internal properties of systems containing satellite plane structures.

In this section, we distinguish between the in-plane and out-of-plane satellite galaxies within the plane structures, and compare their properties. 
{The satellite galaxies are with $\Mstar>10^7\Msun$, similar to that in Section~\ref{satellites}.} 
We calculate the average properties for all in-plane and out-of-plane satellite galaxies in each structure, as indicated by the plus and triangle symbols in Fig.~\ref{satellite_in_out_disc}. 
In general, the properties of in-plane and out-of-plane galaxies  differ significantly. 
{It is found that satellite galaxies within the plane structures tend to be relatively more massive and brighter, while those outside the planes exhibit lower stellar masses and luminosities. 
The difference of metallicity and star formation rate between in-plane and out-of-plane satellites seem weak.
With a careful comparison for the two populations located in the same groups, it is found that most of the in-plane satellites exhibit  higher metallicities and star formation rates compared to the out-of-plane satellites.
However, the color distributions of both in-plane and out-of-plane satellites span a wide range, indicating that there are no clear difference in color index between the two populations.
These findings suggest that in-plane satellites may have slightly longer formation times, and exhibit more active interstellar matter cycles, which could influence the star formation activities, compared to out-of-plane satellites.}

\section{Comparison with Observations}
\label{sect:comparison}

To compare TNG simulation data with observational results, we identify analogs of the Milky Way (MW), Andromeda (M31), and NGC 4490 in the TNG50-1 data, and of the Centaurus A (Cen A) in the TNG100-1 data. 
These analogs are selected primarily based on halo mass.
According to \cite{Patel2017}, the halo mass of MW is approximately $1.5\times10^{12}\Msun$​, and the halo mass of M31 is greater than $1.5\times10^{12}\Msun$.
\cite{Tully2015} estimated that the total mass within the virial radius of Centaurus A is about $8\times10^{12}\Msun$. 
\cite{Pawlowski2024} revised the previously over- and under-estimated halo mass for NGC 4490, concluding that the corrected mass is $2.6\times10^{11}\Msun$. 
{Note that the satellite galaxies samples in TNG50-1 include the galaxies with $\Mstar > 10^5 \Msun$ in this section, due to the comparison with observed satellite plane structures having satellite galaxies with low stellar masses.
However, conclusions for satellite galaxies within $10^5\Msun< \Mstar < 10^7\Msun$ should be explained with caution due to mass resolution limitations.} 

\subsection{Milky Way analogs}
In previous research, \cite{Xu2023} identified Milky Way analogs in the TNG50-1 simulation by restricting the stellar mass of halos to $1\times10^{10}\Msun$​ to $10\times10^{10}\Msun$, setting the stellar mass of satellite galaxies to $\Mstar > 10^5\Msun$​, and requiring a distance from the central galaxy between 15 and 300 $\kpc$. 
They required the groups to contain at least 14 satellite galaxies based on the VPOS and identified 231 MW analogs. 
However, only one halo (Halo ID: 395, $Halo395$) exhibited similarities to the MW in terms of the angle between the satellite plane structure and the central plane, as well as the radial distribution of satellites.
Therefore, we narrow the search range and apply the following selection criteria: the halo mass $\rm M_{200}=0.5\times10^{12}\Msun$ to​ $2.5\times10^{12}\Msun$, satellite galaxy stellar mass $\Mstar > 10^5\Msun$​, and a minimum of 14 satellites within the groups, aiming to distinguish between the three types of satellite structures in MW analogs. 
Based on these criteria, we identify 154 MW analogs. 
The radial distribution of satellites is shown in the panel (a) of Fig.~\ref{comp_1}.
The observational data of MW is taken from \cite{McConnachie2012,Torrealba2016a, Torrealba2019}.

\subsection{M31 analogs}
To explore the characteristics of M31 analogs, \cite{Buck2015} utilized the $N$-body simulation PKDGRAV2 and set the selection criteria as $\rm M_{200}$ from $0.74\times10^{12}\Msun$ to $2.2\times10^{12}\Msun$, satellite masse from $4.4\times10^7\Msun$ to $1.5\times10^{10}\Msun$​​, and distance to the central galaxy from $30 \kpc$ to $250 \kpc$. 
In our research, considering the characteristics of TNG50-1 data, we simplify the selection criteria by setting the mass range from $1\times10^{12}\Msun$ to $3\times10^{12}\Msun$​. 
For the number of satellite galaxies, we use the classical count of 27 satellites within $500 \kpc$ around M31. 
This results in 57 M31 analogs, and we plot their radial satellite distribution in panel (b) of Fig.~\ref{comp_1}.
The observational data of M31 is taken from \cite{Conn2012}.

\subsection{NGC 4490 analogs}
\cite{Pawlowski2024} searched for NGC 4490 analogs in the  IllustrisTNG simulations by selecting systems with masses between $0.2\times10^{12}\Msun$ and $2\times10^{12}\Msun$, distance within $450 \kpc$, and mass less then $2\times10^{11}\Msun$ within $1 \Mpc$. 
They finally identified 141 NGC 4490 analogs. 
Using the random projection fitting method to calculate the system flattening ratio ($D/L$), they found that the peak of sample distribution ($D/L = 0.6$) was significantly larger than the $D/L$ ratio of NGC 4490 ($D/L = 0.38$). 
Therefore, we set the mass range from $0.1\times10^{12}\Msun$ to $0.2\times10^{12}\Msun$, and utilize the NGC 4490 satellite galaxy data taken from \cite{Karachentsev2024}, which provided information on 14 satellite galaxies in NGC 4490 within $450 \kpc$. 
This results 134 NGC 4490 analogs. 
We select the 14 satellites with the largest masses of each analogs to plot their radial distribution, as shown in the panel (c) of Fig.~\ref{comp_1}.

\subsection{Centaurus A analogs}
Using the 27 satellites around Cen A from \cite{Kanehisa2023}, we attempt to find Cen A analogs in TNG50-1 using the previous method, but find few samples that include systems with a satellite plane. 
This is because TNG50-1 contains few dark matter halos as massive as the Cen A halos. 
Therefore, we chose to search Cen A analogs in the TNG100-1 simulation. 
We apply the selection criteria from \cite{Muller2018, Muller2019}, i.e., viral mass $\rm M_{200}=4\times10^{12}\Msun$ to $12\times10^{12}\Msun$, and add the criterion of satellite mass with $\Mstar > 10^{8}\Msun$ and including at least 18 satellites within $500 \kpc$ centred on Cen A. 
Finally, We identify 36 Cen A analogs, and plot their radial satellite distributions, as shown in the panel (d) of Fig.~\ref{comp_1}.

\subsection{radial satellite distributions}

{From Fig.~\ref{comp_1}, it is evident that for local systems like MW and M31, with intermediate stellar masses, the spatial distributions of their satellite galaxies resemble the radial distribution of satellite galaxies in groups with satellite plane structures in TNG50-1.} 
These systems exhibit a dense arrangement within $200 \kpc$ and a more sparse distribution at greater distances. 
In contrast, groups without satellite plane structures show a more uniform satellite distribution, generally appearing more dispersed. 
For the satellites around NGC 4490, their radial distribution appears sparse and uniform. 
Cen A's satellites display a diffuse distribution, primarily concentrated at larger distances. 

\subsection{metallicity -stellar mass and magnitude - color relationships}

Additionally, we compare the properties of MW analogs in TNG50-1 and observational data from \cite{McConnachie2012} on the ``11 classical satellites'' of the Milky Way and CVn I, including the mass, metallicity, $\rm B-V$ color, and V-band absolute magnitude ($\rm M_v$), as well as the mass and metallicity of Antlia II, discovered by \cite{Torrealba2019}.  
We construct the mass-metallicity relation for the Milky Way halo (shown in the left-top panel of Fig.~\ref{comp_2}) and the $\rm B-V$ versus absolute magnitude diagram (shown in the left-bottom panel of Fig.~\ref{comp_2}).
Simultaneously, we use the satellite mass data from \cite{Santos2020} and the metallicity of M31 satellites from \cite{Jennings2023} to compare the mass-metallicity relation of 25 M31 satellites with the TNG50-1 simulation (shown in the right-top panel of Fig.~\ref{comp_2}). 
We also use the $\rm B-V$ colors of M31 satellites from \cite{Conn2012} and the V-band absolute magnitudes from \cite{Savino2022} to construct the corresponding $\rm B-V$ versus absolute magnitude diagram (shown in the right-bottom panel of Fig.~\ref{comp_2}).

From Fig.~\ref{comp_2}, we observe that the masses of Milky Way and Andromeda satellites are generally smaller than the overall masses of plane structures in the simulation.
{The metallicities of satellite galaxies in the MW plane structure are comparable to some of the massive satellites in MW-like plane structures in TNG50-1. 
However, the metallicities of all the satellite galaxies in the M31 plane structure appear to be higher than those in M31-like plane structures in TNG50-1.}
In the bottom panels, it is evident that the satellites in plane structures are distinctly separate from those out-of-plane structures in the magnitude - color digram.
{Moreover, the satellites in the MW and M31 plane structures are located in the B-V color range of 0.0 to 0.2, while the majority of satellites in the plane structures in TNG50-1 exhibit a B-V color index between 0.0 and 0.5. 
This indicates that the satellites in the MW and M31 plane structures are broadly consistent in color index with those in the TNG50-1 plane structures. 
However, the satellites in the MW and M31 plane structures are significantly fainter than their simulated counterparts in TNG50-1. 
As the stellar masses of all the satellites in M31 plane structure are less than $10^7\Msun$, the comparison with the results in TNG50-1 should be carefully treated.}

{The discrepancy between the observational data and the TNG50-1 simulation data may arise from several factors. 
First, galaxy masses in simulations might be overestimated due to the inclusion of intracluster light \citep[ICL; e.g.,][]{Tang2021, Pillepich2018a}. 
Observationally, galaxies are often defined by a surface brightness cutoff, with light fainter than $\rm\sim 26.5\ mag\ arcsec^{-2}$ typically classified as ICL and excluded. 
In contrast, simulations define galaxy structures based on gravitationally bound systems identified by algorithms like SUBFIND, which include ICL. 
This difference can lead to higher mass estimates in simulations compared to observations.
Second, simulations might over-predict galaxy growth due to more frequent mergers and accretion events, which could bias mass and metallicity estimates. 
Finally, the lack of interstellar extinction in simulations may result in satellites appearing brighter than they do in real observations, further contributing to the discrepancy between simulated and observed data.}
\section{Conclusions and Discussion}
\label{sect:conclusion}
In this work, we primarily utilize data from the TNG50-1 simulation at $\rm z=0$ to study the satellite plane structures. 
We calculate their geometric and dynamical properties -- such as system thickness $H$,  scaled system spin $S$, and the number of co-rotating satellites $K$ -- using an improved inertia tensor method. 
{We identify 79 satellite plane structures among 699 samples, most of which are located in the dark matter halos with $\rm 10^{11.5}\Msun<M_{200}<10^{12.5}\Msun$.} 
Most (72) of these plane structures are found to be arranged in a thin, co-rotating plane, with a plane thickness of $5.24\kpc$.
The fraction of satellite planes in this sample is 11.30\%, which is roughly consistent with the initially predicted range of 10\% to 20\%.
To search for Cen A analogs, we use data from TNG100-1. 
Due to resolution limitations in TNG100-1, we increase the satellite mass threshold by a factor of 100.
Using similar methods, we identify 318 satellite plane structures among 1,173 samples, resulting in a fraction of 27.11\%.
{To address the Milky Way satellite plane, we also increase the subhalo selection criterion to N > 14 in TNG50-1. 
The fraction of satellite planes under this condition rises to $\sim30\%$, significantly higher than the original 11.3\% and observational results. 
This is because the number of groups and satellite planes decreases from 699 to 106 and 79 to 32, respectively. 
This finding highlights that selection effects can impact the fraction of satellite planes. 
Nonetheless, diverse satellite plane structures exist in the $\Lambda$CDM universe, similar to observations. 
As \cite{Muller2023} noted: ``Solutions to the plane of satellite problem should therefore not only be tailored to the Milky Way but need to explain all these different observed systems and environments.''} 

{When analyzing the properties of satellite planes in TNG50-1, we categorize the groups into three types: {\it plane, pseudo-plane, and non-plane structures}. 
We examine the stellar mass, luminosity, color, metallicity, and formation time of central and satellite galaxies within these structures. 
We find that the galaxy properties of plane structures different from those of others. 
Central galaxies in plane structures generally have intermediate stellar masses in the range of $10^{10}\Msun$ to $10^{11}\Msun$. 
This suggests that central galaxies with very large or very small masses are less likely to form satellite planes. 
Note that in the intermediate stellar mass range, some central galaxies are also form non-plane structures.
This will be investigated in future analyses. 
Additionally, these central galaxies have low metallicities, but similar  luminosities, star formation rates, color indices, and formation times, compared to those in non-plane structures. 
Satellite galaxies in plane structures are much brighter and bluer than those in non-plane satellites. 
The stellar masses of satellites in the plane structures concentrated  around $\Mstar\sim10^8\Msun$, while those in non-plane structures exhibit a more uniform distribution across stellar mass $\Mstar>10^7\Msun$.
When comparing satellite galaxies within and outside the planes in plane structures, we observe that galaxies within the planes generally have larger masses, higher luminosities, higher metallicities, and greater star formation rates.
There is a weak difference of color indices between in-plane and out-of-plane satellites.}

{Our findings suggest that satellite plane formation is more likely to occur around central galaxies that are star-forming within and fall within intermediate ranges of stellar mass, metallicity, age and luminosity. 
The similarities between satellite galaxies in plane and non-plane  structures for masses $\Mstar>10^{10}\Msun$ infer that these massive satellites may not genuinely belong to the plane structures  but instead appear within the planes by coincidence. 
It is also confirmed that in-plane satellites have slightly longer formation times, and exhibit more active interstellar matter cycles, which could influence the star formation activities.
Future work will explore the evolution of plane and non-plane structures by analyzing merger histories.}

When comparing the simulation results with observational data, we  apply mass and quantity constraints to obtain more samples. 
We find that the radial distribution of satellites around medium-mass groups in TNG50-1 matches observations well, though deviations occur for high-mass or low-mass groups. 
Significant differences are observed in mass, metallicity, $\rm B-V$ color, and magnitude, with simulated plane structures showing higher masses, higher metallicities, higher $\rm B-V$ indices, and greater luminosities than observed satellites in the MW and M31 plane structures. 
These discrepancies likely result from inherent inaccuracies in both observational and simulation data. 
Given the high resolution and small volume of TNG50-1, it is better suited for studying medium to small-mass galaxies, while larger-mass halos may require analysis using TNG100-1.

During this study, we notice some differences of morphological parameters between observed satellite planes and those in simulations. 
Observed satellite planes have an RMS thickness of about $20-50 \kpc$. 
Satellites at greater distances can still be part of the plane, exhibiting coherent motion. 
In contrast, most simulated satellite planes are extremely thin, with most having $H<10\kpc$ in the co-rotating plane formed by the nearest massive satellites. 
For more distant satellites, the coherence of motion and plane thickness are compromised. 
For example, in systems like $Halo395$ from \cite{Xu2023}, nine central satellites nearest the central galaxy are easily identified as co-rotating in the same plane. 
However, when considering the brightest 14 satellites, the main plane of flattening can be completely altered, becoming almost perpendicular to the central co-rotating plane. 
Additionally, in TNG100-1, some satellite plane structures are not as thin as those in TNG50-1, and the velocity vectors of satellites do not always lie in the same plane.

\section*{Acknowledgements}

The authors thank the editors and anonymous referee for the useful suggestions.  
The author thank the Illustris and IllustrisTNG projects for providing simulation data.
L.T. is supported by Natural Science Foundation of Sichuan Province (No. 2022NSFSC1842).  
We also acknowledge support from the NSFC grant (Nos., 12073089, 12273027), the Fundamental Research Funds of China West Normal University (No. 21E029), the Sichuan Youth Science and Technology Innovation Research Team (21CXTD0038), and the National Key Program for Science and Technology Research and Development (No. 2017YFB0203300).

\bibliographystyle{aasjournal}
\bibliography{satellite_plane}
\label{lastpage}

\end{CJK*}
\end{document}